\documentclass[twocolumn,aps,prl,superscriptaddress]{revtex4-1}
\usepackage{amssymb} 
\usepackage{amsmath,bm} 
\usepackage{graphicx} 
\usepackage[normalem]{ulem} 
\usepackage{multirow} 
\usepackage[colorlinks,linkcolor=blue,urlcolor=blue,citecolor=blue]{hyperref} 
\usepackage{lipsum} 
\usepackage[usenames, dvipsnames]{xcolor} 
\usepackage{tensor} 
\usepackage{isotope} 
\usepackage{amsmath} 
\usepackage{float} 

\allowdisplaybreaks[4] 

\begin{document}

\newcommand{\xnw}[1]{{\color{red}#1}}
\renewcommand{\topfraction}{0.85}
\renewcommand{\textfraction}{0.1}
\renewcommand{\floatpagefraction}{0.75}
\newcommand{\bea}{\begin{eqnarray}}
\newcommand{\eea}{\end{eqnarray}}
\newcommand{\nnu}{\nonumber\\}

\title{Unraveling the neutron skin thickness through jet charge in deep inelastic scattering}

\date{\today  \hspace{1ex}}

\author{Shan-Liang Zhang}
\affiliation{State Key Laboratory of Nuclear Physics and Technology, Institute of Quantum Matter, South China Normal University, Guangzhou 510006, China}
\affiliation{Department of Physics, Hubei University, Wuhan 430062, China}

\author{Enke Wang}
\email{wangek@scnu.edu.cn}
\affiliation{State Key Laboratory of Nuclear Physics and Technology, Institute of Quantum Matter, South China Normal University, Guangzhou 510006, China}
\affiliation{Guangdong Basic Research Center of Excellence for Structure and Fundamental Interactions of Matter, Guangdong Provincial Key Laboratory of Nuclear Science, Guangzhou 510006, China}

\author{Xin-Nian Wang}
\email{xnwang@ccnu.edu.cn}
\affiliation{Key Laboratory of Quark and Lepton Physics (MOE) $\&$ Institute of Particle Physics,
Central China Normal University, Wuhan 430079, China}

\author{Hongxi Xing}
\email{hxing@m.scnu.edu.cn}
\affiliation{State Key Laboratory of Nuclear Physics and Technology, Institute of Quantum Matter,
South China Normal University, Guangzhou 510006, China}
\affiliation{Guangdong Basic Research Center of Excellence for Structure and Fundamental Interactions of Matter, Guangdong Provincial Key Laboratory of Nuclear Science, Guangzhou 510006, China}
\affiliation{Southern Center for Nuclear-Science Theory (SCNT), Institute of Modern Physics, Chinese Academy of Sciences, Huizhou 516000, China}

\begin{abstract}
The neutron skin thickness in neutron-rich nuclei has traditionally been measured using elastic fixed target electron-nucleus scattering since the 1970s. In this paper, we propose a novel probe of the neutron skin thickness through deep inelastic scattering in electron-ion collisions, leveraging the intrinsic correlation between final-state jet charge distribution and initial-state partonic distributions in nucleons. Specifically, we demonstrate the sensitivity of jet charge distribution to the neutron skin thickness in $e$+Pb collisions with varying centralities, and in isobar collisions of $e$+Ru and $e$+Zr. We predict a strong suppression of positive jet charge distribution and an enhancement for negative jet charge distribution in peripheral electron-ion collisions, revealing the neutron skin effect. This proposal can also be extended to photon and Z-boson tagged jet charge distribution in proton-nucleus collisions at the Large Hadron Collider, providing an alternative access to neutron skin thickness.
 
\end{abstract}


\maketitle

\vspace{0.5cm}
\textbf{\emph{ Introduction}}. 
Understanding the density distribution of protons and neutrons, particularly the neutron skin thickness in nuclei, is crucial for constraining the fundamental properties of cold nuclear matter and the equation of state of neutron star \cite{Roy:2024fxy,Sammarruca:2023mxp,Ma:2022dbh}. Since the 1970s \cite{Frois:1977hr}, the neutron skin thickness, $\Delta R_{np}=R_n-R_p$, defined as the difference in root mean square (rms) radii between neutrons ($R_n$) and protons ($R_p$) in nuclei, has been primarily measured using elastic fixed-target electron-nucleus scattering~\cite{Clark:2002se,Friedman:2012pa,Tarbert:2013jze,Brown:2007zzc,Roca-Maza:2015eza}. While proton radii have been accurately determined for many nuclei using electromagnetic scattering, determining neutron density distribution is more challenge due to nonperturbative uncertainties in hadronic probes and the low scattering rate for parity-violating electron scatterings. These challenges make it difficult to precisely constrain the neutron skin thickness. 

Among various neutron rich nuclei, the doubly magic nucleus $^{208}$Pb is of particular interest due to its significant neutron excess. Recently, the PREX-II collaboration updated the measurement of neutron skin thickness for $^{208}$Pb, reporting $\Delta R^\text{Pb}_{np}=0.283 \pm 0.071$ fm \cite{PREX:2021umo}, significantly reducing the uncertainties compared to previous results $\Delta R^\text{Pb}_{np}=0.33_{-0.18}^{+0.16}$ fm~\cite{Abrahamyan:2012gp}. This measurement, derived from parity-violating asymmetries in polarized electron scattering, reveals some tension with ab initial calculations using chiral effective theory \cite{Hu:2021trw} and other indirect measurements using hadronic probes and astrophysical observations~\cite{Reed:2021nqk}. For instance, coherent $\pi$ photoproduction measurements reported $\Delta R^\text{Pb}_{np}$= 0.15$\pm$0.03 fm~\cite{Tarbert:2013jze}. In addition to these measurements using low energy facilities, new approaches have recently been proposed to extract neutron skin thickness from measurements in relativistic heavy-ion collisions~\cite{Li:2019kkh,Liu:2023pav,Guo:2023nmm,Liu:2023qeq,Cheng:2023ucp,Liu:2022xlm,Liu:2022kvz,Xu:2021uar,Xu:2021vpn}. In particular, through a global analysis within the hydrodynamic model of heavy-ion collisions, a neutron skin of $\Delta R^\text{Pb}_{np}$= 0.217$\pm$0.058 fm was obtained for $^{208}$Pb \cite{Giacalone:2023cet}, competitive with PREX-II measurement. Likewise, taking advantages of current programs at Relativisitic Heavy Ion Collider (RHIC) and the Large Hadron Collider (LHC), hard probes, such as the ratio of $W^+/W^-$~\cite{Paukkunen:2015bwa} and $h^{-}/h^{+}$~\cite{Helenius:2016dsk,ATLAS:2019ibd}, as well as charged particles in isobar collisions~\cite{vanderSchee:2023uii}, are also proposed to provide some new insights into the nuclear structures.           

\begin{figure}[!t]
  \centering
\includegraphics[width=0.4\textwidth]{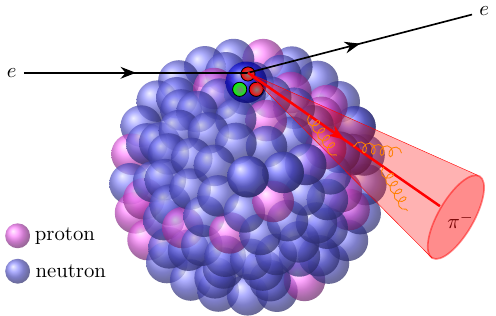}
  \caption{(Color online) An illustration depicting negative jet charge production in  $e$+A collisions on a neutron-rich surface.   
  }\label{eA_skin}
\end{figure}

To achieve a consensus on the quantitative constraints of neutron skin thickness, it is crucial to employ a clean probe with minimal model dependence. In this paper, we propose a novel probe for neutron skin thickness through deep inelastic scattering (DIS) in electron-ion collisions. Unlike the conventional approach using elastic electron-nucleus scattering at low energies, where the nucleus remains intact, the breaking of the nucleus in our new proposal allows us to investigate the nuclear structure using hard probes at partonic level and leverage perturbative QCD for high precision. In terms of partonic structure, the primary difference between proton and neutron lies in the distribution of u-quark and d-quark within them, which is highly correlated with the final-state flavor distributions in neutral-current DIS process. This intrinsic correlation between the final-state flavor tagging and the initial-state u- and d-quark distributions, in turn, reveals the proton and neutron density distributions within the nucleus, see Fig. \ref{eA_skin}. 

We propose using the jet charge distribution as a proxy to u- and d-quark initiated jet, which has been shown to be a useful probe for nucleon structure \cite{Krohn:2012fg,Kang:2020fka} and extensively studied at RHIC and the LHC ~\cite{ATLAS:2015rlw,CMS:2017yer,CMS:2020plq,Chen:2019gqo}. We demonstrate for the first time the sensitivity of this observable to the neutron skin of the nucleus at the upcoming Electron-Ion Collider (EIC)~\cite{AbdulKhalek:2021gbh} with varying centralities. Our findings indicate that the neutron skin results in the increase of negatively charged jets and decrease of positively charged jets in the final state, particularly in peripheral collisions. This method enhances the EIC's capability to study nuclear structure at the nucleon level, linking physics phenomena at both low and high collision energies. 

\textbf{\emph{Jet Charge and neutron skin.}}
The concept of jet electric charge was initially proposed by Feynman and Field~\cite{Field:1977fa}. It can be constructed from the definition~\cite{Krohn:2012fg,Waalewijn:2012sv}
\bea
Q_J=\sum_{h\in {\rm jet}}\left(\frac{p_T^h}{p_T^J}\right)^\kappa Q_h ,
\eea
where $p_T^h$ and $Q_h$ are the transverse momentum and electric charge, respectively, of the identified hadron within the jet with transverse momentum $p_T^J$, $\kappa$ is a free parameter that controls the shapes of the jet charge distribution. 

In the center-of-mass frame of electron-nucleus deep inelastic scattering, $e+{\rm A} \to e + {\rm jet} + X$, the jet charge distribution in the back-to-back region of the outgoing electron-jet can be derived within  the transverse momentum dependent (TMD) QCD factorization ~\cite{Liu:2018trl,Kang:2020fka}
\bea \label{eq-fac}
    \frac{d\sigma}{dQ_J} = & \sum_i e_i^2 \mathcal{F.T.} \bigg\{\tilde f_{i/A}(x,b_T) S_J(b_T,R)\nnu
    &\times H_{ei\to ei}(Q) \mathcal{G}_i(Q_{J},p_T^J R)\bigg\},
\eea
where $\mathcal{F.T.}$ stands for the Fourier transform from coordinate space to momentum space, $\tilde f_{i/A}(x,b_T)$ is the nuclear TMD distribution function \cite{Alrashed:2021csd,Alrashed:2023xsv}, $S_J$ is the soft function, and $H$ is the hard factor with $Q$ representing the invariant mass of the virtual photon in neutral-current DIS process. $\mathcal{G}_i$ is the fragmenting jet function with identified jet electric charge $Q_J$. The scale dependence in these functions is implicit.

As shown in Eq. (\ref{eq-fac}), the initial parton $i$ from the incident nucleus involved in the hard interaction is correlated with the flavor of final observed jet, making the jet flavor tagging an excellent probe of nuclear partonic structure. Noticing the primary difference in the flavor structure between proton and neutron lies in the distribution of u-quark and d-quark. Therefore, identifying u-quark and d-quark tagged jets in the final state can reveal the proton and neutron density distributions within the nucleus. 
We will utilize PYTHIA event generator, which has been shown to be consistent with the TMD factorization results \cite{Kang:2020fka}, to simulate the jet charge distribution in DIS and demonstrate this correlation.

In our simulations, we investigate the neutron skin effect by examining the centrality-dependent jet charge distribution in $e$A collisions,  
\bea
\frac{d\sigma}{dQ_J}&= 2\pi \sum_i\int_{b_{min}}^{b_{max}} dbb f_{i/A}(x,\mu;b)\otimes \frac{d\sigma^{i}}{dQ_J},
\eea
where $b$ is the impact parameter, and the centrality class is determined by $b_{min}$ and $b_{max}$ \cite{Miller:2007ri}. The centrality dependence is encoded into the impact parameter dependent nuclear parton distribution functions (nPDFs). We use the following effective factorization form to parametrize the impact-parameter dependent nPDFs,
\bea
\label{eq:taa}
f_{i/A}(x,\mu;b)&=\frac{T_A^p(b)}{T_A(b)}f_{i}^{p/A}(x,\mu)+ \frac{T_A^n(b)}{T_A(b)}f_i^{n/A}(x,\mu),
\eea
where $T_A (b)=\int dz \rho_A(r)$ is the thickness function of nucleus A, $T_A^p(b)$ and $T_A^n(b)$ are the corresponding thickness functions for proton and neutron, respectively.

We implement the flavor-dependent nuclear distribution using the deformed Woods-Saxon (WS) distribution, 
\bea
\rho_{n,p}(r)=\frac{\rho^0_{n,p}}{1+\exp\left(\frac{r-c_{n,p}}{a_{n,p}}\right)},
\eea
where $\rho^0_{n,p}$ are the saturation densities, $c_{n,p}$ and $a_{n,p}$ represent the half-density radius and skin thickness, respectively. The total nuclear density is $\rho_A(r)=\rho_n(r)+\rho_p(r)$. Following the strategy in determining the neutron skin of Pb from LHC data \cite{Giacalone:2023cet}, we fix the WS parameters for the proton as $c_p$= 6.680 fm and $a_p$=0.448 fm, corresponding to an rms proton radius in Pb $R_p$= 5.436 fm. The parameter $c_n$= 6.690 fm is fixed by the experimental constrain of interpreting $\Delta R_{np}$ as an increase of the neutron surface diffuseness rather than an increase of the neutron half-density radius~\cite{Trzcinska:2001sy,Zenihiro:2010zz}. The neutron skin thickness parameter $a_n$ is determined from different values of $\Delta R_{np}$ from the relation $\langle r^2\rangle_n   \approx \frac{3}{5}c_n^2 +\frac{7}{5} \pi^2a_n^2$ \cite{Trzcinska:2001sy}.   
\begin{table}[t]
\caption{WS parameters and neutron skin thickness for $_{82}^{208}$Pb, $_{44}^{96}$Ru and $_{40}^{96}$Zr used in our numerical calculations.} 
\label{tab:WSparameters}
\begin{tabular}{cccccc}
\hline
nucleus & $c_p\,$[fm] & $a_p\,$[fm] & $c_n\,$[fm] & $a_n\,$[fm] & $\Delta R_{np}$ \\
\hline
$_{82}^{208}$Pb \cite{Tarbert:2013jze} & 6.68 & 0.448 & 6.69 & 0.448 & 0 \\
$_{82}^{208}$Pb \cite{Tarbert:2013jze} & 6.68 & 0.448 & 6.69 & 0.566$^{+0.028}_{-0.045}$ & 0.15$\pm$ 0.03 \\
$_{82}^{208}$Pb \cite{PREX:2021umo} & 6.68 & 0.448 & 6.69 & 0.654$^{+0.044}_{-0.046}$ & 0.28$\pm$ 0.071 \\
$_{44}^{96}$Ru \cite{Xu:2021uar} & 5.06 & 0.493 & 5.075 & 0.505 & 0.03 \\
$_{40}^{96}$Zr \cite{Xu:2021uar} & 4.915 & 0.521 & 5.015 & 0.574 &  0.16 \\
\hline
\end{tabular}
\end{table}

\textbf{\emph{Phenomenological results.}} We now present phenomenological results to demonstrate the sensitivity of nuclear skin on jet charge distribution at the EIC. The simulation is carried out using PYTHIA 8.3 \cite{Sjostrand:2007gs} for $e$+A collisions at $\sqrt{s}=105$ GeV. We implement the default CT14nlo~\cite{Dulat:2015mca} and EPPS16~\cite {Eskola:2016oht} parameterization for nuclear parton distribution functions in $e$+A collisions in PYTHIA 8.3. DIS events are selected by requiring the invariant mass of the exchanged virtual photon $Q^2 \ge 10 $ GeV$^2$. Both the electron's and jet's transverse momentum are selected within the phase space $p_T^{e,J}\ge 10$ GeV and rapidity $|\eta^{e,J}|\le 2$. Jets are reconstructed with the FASTJET anti-$k_t$ jet algorithm \cite{Cacciari:2008gp,Cacciari:2011ma}, with a radius parameter $R=1.0$. We chose $\kappa=0.3$ for jet charge distribution, as used in Ref.~\cite{Kang:2020fka}, which allows for effective jet flavor identification~\cite{Kang:2021ryr,Kang:2023ptt}. For the simplicity of illustration, we consider $\pi^\pm$ distribution in jet, as they play a dominant role in jet charge distribution.

In this study, we use three sets of neutron
skin thickness for $^{208}$Pb to illustrate the sensitivity of jet charge distribution to the neutron skin and demonstrate the differential power of jet charge distribution on the current knowledge of neutron skin thickness: 1. $\Delta R_{np}^{\rm Pb}$= 0 fm representing a vanishing neutron skin; 2. $\Delta R_{np}^{\rm Pb}$= 0.15$\pm$0.03 fm, taken from coherent $\pi$ photoproduction measurements~\cite{Tarbert:2013jze}; 3. $\Delta R_{np}^{\rm Pb}$= 0.28$\pm$0.071 fm, taken from the PREX-II experiment~\cite{PREX:2021umo}. The exact values of the WS parameters for these sets and for isobar Ru and Zr are listed in Tab. \ref{tab:WSparameters}.

\begin{figure}[!t]
  \centering
\includegraphics[width=0.45\textwidth]{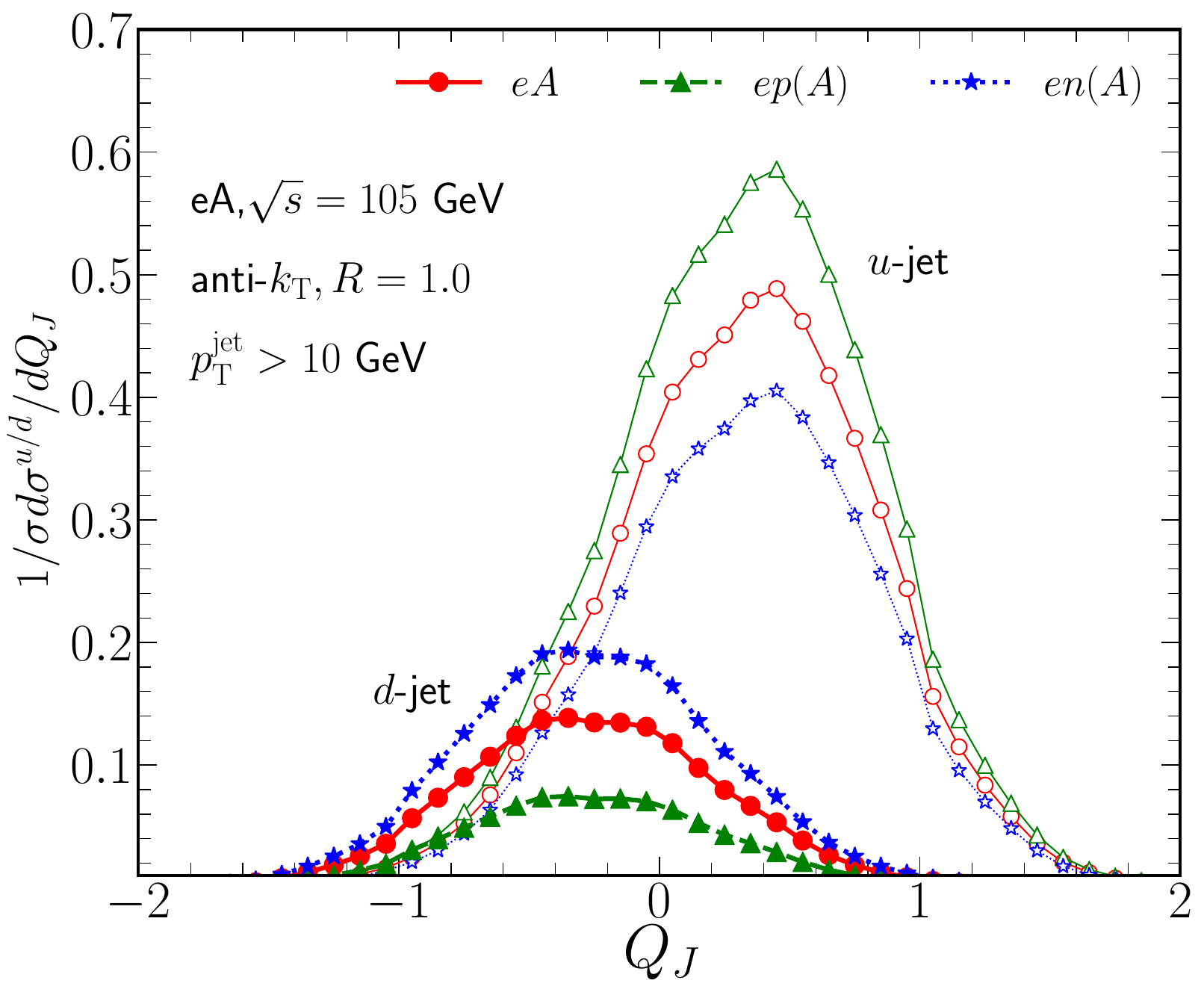}
  \caption{(Color online) Jet charge distribution of $u$ (hollow) and $d$ (filled)  quark initiated-jet in ep(A) (Triangle), en(A) (Star) and eA (Circle) collisions, normalized by the total jet cross section.   
  }\label{jetcharge_dis}
\end{figure}
 
To illustrate the discriminative power of the jet charge distribution on u- and d-quark jet, we plot in Fig. \ref{jetcharge_dis} the u-quark (hollow) and d-quark (filled) quark initialed-jet distribution as a function of jet charge $Q_J$, normalized by the corresponding total jet cross section. The contributions from the bounded proton (denoted as ``$ep(A)$") and neutron (denoted as ``$en(A)$") in Pb are represented, respectively, by triangle and star markers, while the total contribution from $e$+Pb collisions are shown in circle marker. In this illustration, the neutron skin thickness is set to 0, and the flavor of the final jet is tagged by the parton produced from the electron-parton hard scattering. As we expected, the positive jet charge is dominated by the u-quark channel, while d-quark channel mainly contribute in the negative jet charge region. These results indicate that the u-quark and d-quark jet can be well separated within different jet charge regions. It is important to note that u-quark jets play a major role across a wide range of $Q_J$, primarily due to the factor of four stronger electromagnetic coupling strength between the virtual photon and the u-quark comparing to that for the d-quark. This dominant role of u-quark is even more pronounced for the contribution from protons bounded in Pb as shown by the green lines.

\begin{figure*}[ht]
  \centering
\includegraphics[width=1.0\textwidth]{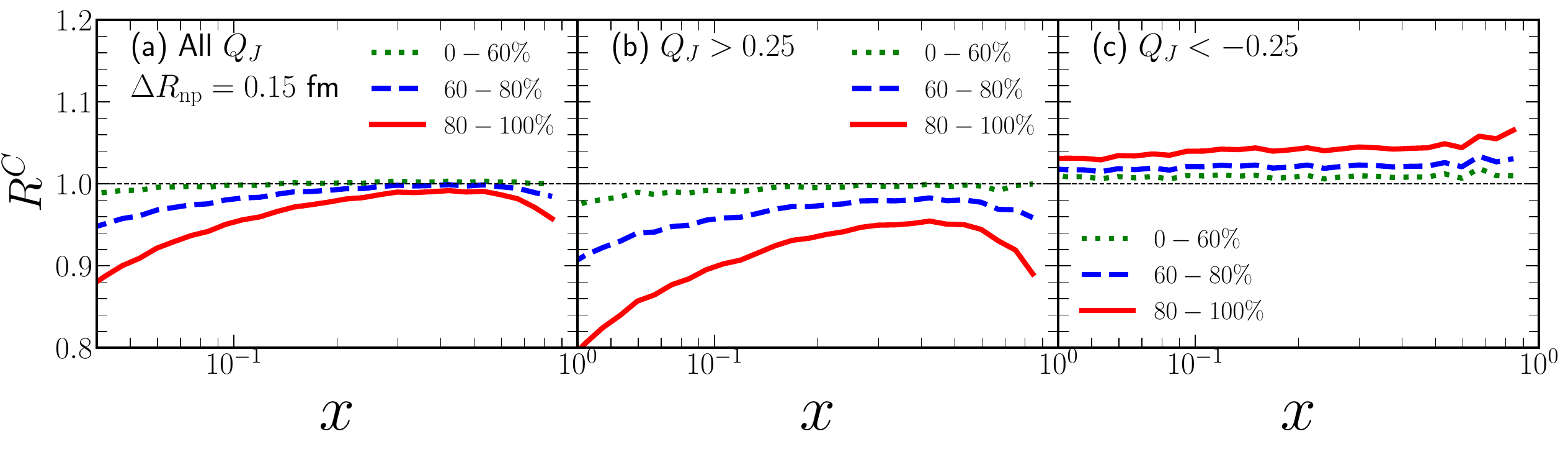}
  \caption{(Color online)  The nuclear modification factor $R^C$ in 0-60$\%$, 60-80$\%$ and 80-100$\%$ centrality intervals with (a) all jet charge $Q_J$, (b) $Q_J>0.25$, (c) $Q_J<-0.25$.     
  }\label{ns-jetchgarge}
\end{figure*}

According to the above illustration, one would expect significant reduction in the positive jet charge distribution if the virtual photon interacts with the neutron skin region, due to the relatively lower u-quark compositions in neutrons compared to protons. This phenomenon can be effectively examined as a function of centrality in $e$+A collisions, which serves as a proxy for the spatial position of the struck nucleon within the nucleus. To quantify the correlation between nuclear skin and the jet charge distribution as a function of centrality, we propose measuring the centrality-dependent nuclear modification factor $R^{C}$. This factor is defined as the ratio of the jet charge distribution in a given centrality to that in minimum-bias (MB) events,
\bea
R^{C}(x,Q_J)=\frac{d\sigma^\text{C}/dxdQ_J}{d\sigma^\text{MB}/dxdQ_J},
\label{rcp}
\eea
where $x$ is the momentum fraction of parton $i$ in nucleus per nucleon at leading order kinematics, defined as $x=\frac{e^{\eta^J}p_T^J}{E_e-e^{-\eta^J}p_T^J}$~\cite{Kang:2013wca}, with $E_e$ being the incoming electron energy in the center of mass frame. In this observable, we expect a significant cancellation of the uncertainties from nPDFs in the ratio and strong sensitivities to the proton and neutron spatial densities encoded in the thickness functions in Eq. (\ref{eq:taa}).

Taking a moderate neutron skin thickness as an example, i.e. $\Delta R_{np}^{\rm Pb} = 0.15$ fm, we show in Fig.~\ref{ns-jetchgarge} the numerical results for $R^{C}$ in 0-60$\%$, 60-80$\%$ and 80-100$\%$ centrality intervals with (a) all jet charge $Q_J$ , (b) $Q_J>0.25$, (c) $Q_J<-0.25$. As can be seen, even without specific jet charge selection, $R^{C}$ exhibits a moderate sensitivity to centralities, which is caused by the dominance of u-quark jets in DIS as shown in Fig. \ref{jetcharge_dis}. The suppression of $R^{C}$ reveals the reduction of initial u-quark contribution in the presence of neutron skin in peripheral collisions, while such suppression is negligible in central collisions due to the absence of neutron skin effect in the center of the nucleus. This sensitivity is further enhanced when a selection cut $Q_J>0.25$ is applied, as the role of u-quark jet becomes more significant in this region. On the contrary, the increase of d-quark jets and suppression of u-quark jets in $Q_J<-0.25$ region lead to the enhancements of $R^{C}$ with increasing centralities.

Though the uncertainty of nPDFs is expected to be largely cancelled in the ratio defined in Eq. (\ref{rcp}), the bias from classification of event geometry, as specified by the centrality assignment, could pose a challenge in DIS process. To further mitigate such potential ambiguity, we propose measuring the following double ratio
\bea
\mathbb{R} &\equiv& \frac{R^{C}(x,Q_J\le -Q_c)}{R^{C}(x,Q_J\ge Q_c)} 
\nnu
&=& \frac{d\sigma^\text{C}/dxdQ_J|_{Q_J\le -Q_c}}{d\sigma^\text{C}/dxdQ_J|_{Q_J\ge Q_c}}
\frac{d\sigma^\text{MB}/dxdQ_J|_{Q_J\ge Q_c}}{d\sigma^\text{MB}/dxdQ_J|_{Q_J\le -Q_c}},
\label{eq-doubleR}
\eea
where $Q_c$ is a parameter for specific selection of jet charge regions and set to 0.25 in our simulations. In Eq. (\ref{eq-doubleR}), the centrality selection bias should be significantly reduced by the first ratio. Additionally, we expect the uncertainty from the initial-state nPDFs to become insignificant compared to that arising from the neutron skin thickness if the nPDF uncertainty is significantly reduced through high-precision measurements in the current proton-nucleus program at the LHC and in future EIC experiments \cite{AbdulKhalek:2021gbh,Anderle:2021wcy}.
Moreover, another possible contamination is the effect of final state multiple scattering, such as the transverse momentum broadening \cite{Guo:1998rd,Kang:2013raa,Ru:2019qvz,Ru:2023ars} and parton energy loss \cite{Guo:2000nz,Wang:2002ri,Li:2020rqj} in cold nuclear matter, which also lead to nuclear modification of the final jet charge distribution. However, such an effect is driven by the strong interaction of QCD and is blind to the electric charge. Therefore, it is expected that these final-state effects are similar for u-quark and d-quark jets and will also be significantly cancelled in the double ratio $\mathbb{R}$.    

\begin{figure}[!t]
  \centering
\includegraphics[width=0.45\textwidth]{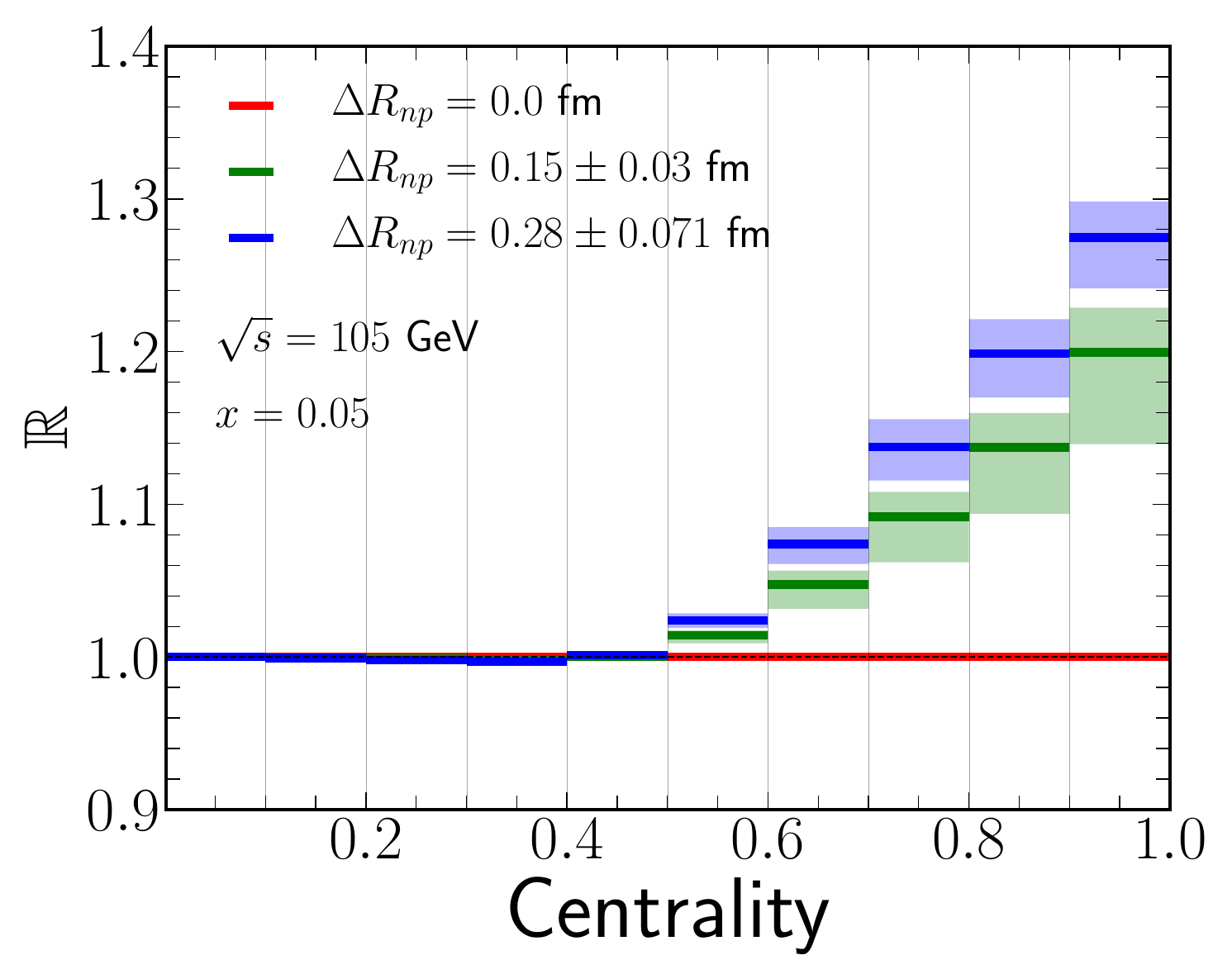}
  \caption{(Color online)  The double ratio of $R^{CP}$ at $x=0.05$ for $Q_J<-0.25$  to  $Q_J>0.25$ in $e$+Pb collisons with the halo-type neutron-skin thicknesses of 0.0 fm, 0.15 fm and 0.28 fm respectively. 
  }\label{skin_C}
\end{figure}

Shown in Fig.~\ref{skin_C} are the results for $\mathbb{R}$ evaluated at $x=0.05$ with different sets of neutron-skin thicknesses. The bands represent the uncertainties of the extracted $\Delta R_{np}^{\rm Pb}$ from experimental data~\cite{PREX:2021umo,Tarbert:2013jze}. With larger neutron-skin thicknesses for Pb, we observe greater deviations from unity (representing non-neutron skin effect) in peripheral collisions, reaching up to $30\%$ in the most peripheral 90-100$\%$ $e$+Pb collisions. Notably, the two sets of $\Delta R_{np}^{\rm Pb}$ extracted from existing experimental measurements can be well distinguished by this novel observable, taking full advantage of jet charge. Considering the high luminosity design and high statistics of the EIC ~\cite{AbdulKhalek:2021gbh}, the uncertainties from systematic and statistical errors are expected to be much smaller than the bands shown in Fig.~\ref{skin_C}, promising accurate determination of neutron skin thickness at the future EIC. 

\begin{figure}[!t]
  \centering
\includegraphics[width=0.45\textwidth]{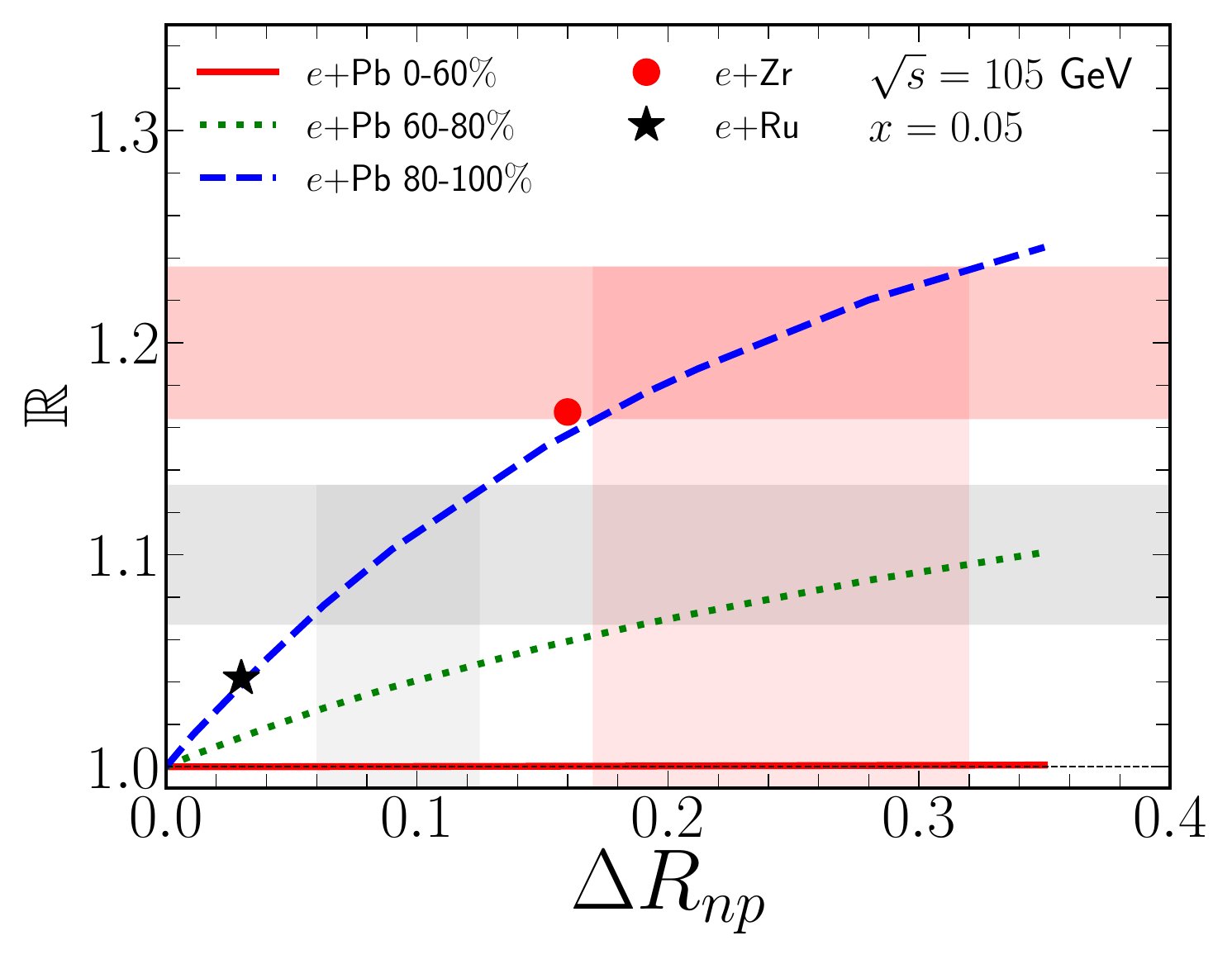}
\caption{(Color online)  The double ratio $\mathbb{R}$  evaluated as a function of $\Delta R_{np}$ in three centralities of $e$+Pb (lines) collisions, 80-100$\%$ $e$+Ru (solid circle) and $e$+Zr (solid star) collisions. The vertical bands show the variation of $\mathbb{R}$ due to uncertainties of $\Delta R_{np}$, while the horizontal bands represent the impact of $3\%$ uncertainty of $\mathbb{R}$ on $\Delta R_{np}$.
  }\label{skin_C_x}
\end{figure}

Shown in Fig.~\ref{skin_C_x} are results for $\mathbb{R}$ as a function of neutron-skin thicknesses of $^{208}$Pb, $_{44}^{96}$Ru and $_{40}^{96}$Zr. As we observe, $\mathbb{R}$ for $e$+Pb with different centralities, shown by the curves, increases monotonically with neutron-skin thickness, and the growth rate rises with centrality. The vertical bands show the variation of $\mathbb{R}$ due to uncertainties in the extracted neutron skin thickness on $\mathbb{R}$, while the horizontal bands represent the impact of $3\%$ experimental uncertainty of $\mathbb{R}$ on the extraction of  neutron skin thickness. The predicted sensitivity to the relative differences in the small-neutron-skin thickness region is particularly impactful for probing the differences between $_{44}^{96}$Ru and $_{40}^{96}$Zr. For illustration, we use the WS parameters for $_{44}^{96}$Ru and $_{40}^{96}$Zr from Ref.~\cite{Xu:2021uar}, as listed in Table \ref{tab:WSparameters}. The small neutron skin thickness of $_{44}^{96}$Ru results in small impact on the double ratio in peripheral collisions (80-100$\%$) as shown by the black star marker, while $_{40}^{96}$Zr, shown by red circle point result in a $17\%$ change in the double ratio in peripheral collisions. 

\textbf{\emph{Summary and discussions.}} Taking full advantage of the intrinsic correlation between final-state jet flavor information and initial-state nuclear partonic structure in high-energy deep inelastic electron-nucleus collisions, we propose using jet charge distribution as
a novel hard probe to measure neutron skin thickness, differing from the traditionally method of low energy elastic fixed target electron-nucleus scattering. Taking $e$+Pb collision as an example, we demonstrate the sensitivity of jet charge distributions in peripheral collisions to the neutron skin effect at the EIC. We calculated the peripheral to minimum-bias event ratio $R^{C}$ and predicted strong suppression in positive jet charge region and enhancement in negative jet charge region, revealing the impact of neutron skin thickness on the u-quark and d-quark jet production in peripheral $e$+A collisions. Furthermore, through the double ratio of $\mathbb{R}$, we expect improved sensitivity to neutron skin thickness and mitigation of bias from centrality classification, as well as final-state cold nuclear matter effects. Specifically, considering the WS parameters from the state-of-art measurement of PREX-II, we observe up to 30$\%$ deviations from unity, representing non-neutron skin effects, in the most peripheral 90-100$\%$ $e$+Pb collisions. We also observe pronounced sensitivity to the relative differences in the small-neutron-skin thickness region, leading to more than $10\%$ change in the double ratio in peripheral collisions between $e$+Ru and $e$+Zr collisions. Our proposal extends the current programs of EIC, providing a novel method to accurately probe the neutron skin effect. With future high luminosity and statistics, more accurate determination of the neutron skin will be feasible.

We emphasize that our proposal of using jet charge to probe neutron skin thickness is not limited to the EIC and can have broad applications. For example, the idea can be extended to jet charge distribution in p+Pb program at the LHC. It has been show that prompt photon and Z-boson tagged jets are dominated by quark jets~\cite{Zhang:2021oki}, providing another golden channel to access the initial u-quark and d-quark distributions. This intrinsic correlation offers an alternative approach to measure neutron skin thickness. Detailed phenomenological investigations will be presented in a future publication. Furthermore, joint distribution of jet charge and particle multiplicity can carry more information for u- and d-quark jet discrimination~\cite{Kang:2023ptt} at the LHC, which can be implemented to further improve the sensitivity of jet charge on neutron skin effect.  

{\bf Acknowledgments:} 
We thank Weiyao Ke and Xiaohui Liu for useful discussions. This research is supported by the National Natural Science Foundation of China with Project No. 12405152, 12035007, 12475139, 11935007.

\vspace*{-.6cm}

\begin{thebibliography}{10}

\bibitem{Roy:2024fxy}
P.~Roy and R.~Kanungo,
\newblock DAE Symp. Nucl. Phys. {\bf 67}, 23 (2024).

\bibitem{Sammarruca:2023mxp}
F.~Sammarruca,
\newblock Symmetry {\bf 16}, 34 (2024), arXiv:2311.02539.

\bibitem{Ma:2022dbh}
Y.-G. Ma and S.~Zhang,
\newblock {\em {Influence of Nuclear Structure in Relativistic Heavy-Ion
  Collisions}} (, 2022), pp. 1--30, arXiv:2206.08218.

\bibitem{Frois:1977hr}
B.~Frois {\em et~al.},
\newblock Phys. Rev. Lett. {\bf 38}, 152 (1977).

\bibitem{Clark:2002se}
B.~C. Clark, L.~J. Kerr, and S.~Hama,
\newblock Phys. Rev. C {\bf 67}, 054605 (2003), arXiv:nucl-th/0209052.

\bibitem{Friedman:2012pa}
E.~Friedman,
\newblock Nucl. Phys. A {\bf 896}, 46 (2012), arXiv:1209.6168.

\bibitem{Tarbert:2013jze}
C.~M. Tarbert {\em et~al.},
\newblock Phys. Rev. Lett. {\bf 112}, 242502 (2014), arXiv:1311.0168.

\bibitem{Brown:2007zzc}
B.~A. Brown, G.~Shen, G.~C. Hillhouse, J.~Meng, and A.~Trzcinska,
\newblock Phys. Rev. C {\bf 76}, 034305 (2007).

\bibitem{Roca-Maza:2015eza}
X.~Roca-Maza {\em et~al.},
\newblock Phys. Rev. C {\bf 92}, 064304 (2015), arXiv:1510.01874.

\bibitem{PREX:2021umo}
PREX, D.~Adhikari {\em et~al.},
\newblock Phys. Rev. Lett. {\bf 126}, 172502 (2021), arXiv:2102.10767.

\bibitem{Abrahamyan:2012gp}
S.~Abrahamyan {\em et~al.},
\newblock Phys. Rev. Lett. {\bf 108}, 112502 (2012), arXiv:1201.2568.

\bibitem{Hu:2021trw}
B.~Hu {\em et~al.},
\newblock Nature Phys. {\bf 18}, 1196 (2022), arXiv:2112.01125.

\bibitem{Reed:2021nqk}
B.~T. Reed, F.~J. Fattoyev, C.~J. Horowitz, and J.~Piekarewicz,
\newblock Phys. Rev. Lett. {\bf 126}, 172503 (2021), arXiv:2101.03193.

\bibitem{Li:2019kkh}
H.~Li {\em et~al.},
\newblock Phys. Rev. Lett. {\bf 125}, 222301 (2020), arXiv:1910.06170.

\bibitem{Liu:2023pav}
Q.~Liu, S.~Zhao, H.-j. Xu, and H.~Song,
\newblock Phys. Rev. C {\bf 109}, 034912 (2024), arXiv:2311.01747.

\bibitem{Guo:2023nmm}
W.-M. Guo, B.-A. Li, and G.-C. Yong,
\newblock Phys. Rev. C {\bf 108}, 034617 (2023), arXiv:2307.05135.

\bibitem{Liu:2023qeq}
L.-M. Liu, J.~Xu, and G.-X. Peng,
\newblock Phys. Lett. B {\bf 838}, 137701 (2023), arXiv:2301.07893.

\bibitem{Cheng:2023ucp}
Y.-L. Cheng, S.~Shi, Y.-G. Ma, H.~St\"ocker, and K.~Zhou,
\newblock Phys. Rev. C {\bf 107}, 064909 (2023), arXiv:2301.03910.

\bibitem{Liu:2022xlm}
L.-M. Liu, C.-J. Zhang, J.~Xu, J.~Jia, and G.-X. Peng,
\newblock Phys. Rev. C {\bf 106}, 034913 (2022), arXiv:2209.03106.

\bibitem{Liu:2022kvz}
L.-M. Liu {\em et~al.},
\newblock Phys. Lett. B {\bf 834}, 137441 (2022), arXiv:2203.09924.

\bibitem{Xu:2021uar}
H.-j. Xu {\em et~al.},
\newblock Phys. Rev. C {\bf 108}, L011902 (2023), arXiv:2111.14812.

\bibitem{Xu:2021vpn}
H.-j. Xu, H.~Li, X.~Wang, C.~Shen, and F.~Wang,
\newblock Phys. Lett. B {\bf 819}, 136453 (2021), arXiv:2103.05595.

\bibitem{Giacalone:2023cet}
G.~Giacalone, G.~Nijs, and W.~van~der Schee,
\newblock Phys. Rev. Lett. {\bf 131}, 202302 (2023), arXiv:2305.00015.

\bibitem{Paukkunen:2015bwa}
H.~Paukkunen,
\newblock Phys. Lett. B {\bf 745}, 73 (2015), arXiv:1503.02448.

\bibitem{Helenius:2016dsk}
I.~Helenius, H.~Paukkunen, and K.~J. Eskola,
\newblock Eur. Phys. J. C {\bf 77}, 148 (2017), arXiv:1606.06910.

\bibitem{ATLAS:2019ibd}
ATLAS, G.~Aad {\em et~al.},
\newblock Eur. Phys. J. C {\bf 79}, 935 (2019), arXiv:1907.10414.

\bibitem{vanderSchee:2023uii}
W.~van~der Schee, Y.-J. Lee, G.~Nijs, and Y.~Chen,
\newblock (2023), arXiv:2307.11836.

\bibitem{Krohn:2012fg}
D.~Krohn, M.~D. Schwartz, T.~Lin, and W.~J. Waalewijn,
\newblock Phys. Rev. Lett. {\bf 110}, 212001 (2013), arXiv:1209.2421.

\bibitem{Kang:2020fka}
Z.-B. Kang, X.~Liu, S.~Mantry, and D.~Y. Shao,
\newblock Phys. Rev. Lett. {\bf 125}, 242003 (2020), arXiv:2008.00655.

\bibitem{ATLAS:2015rlw}
ATLAS, G.~Aad {\em et~al.},
\newblock Phys. Rev. D {\bf 93}, 052003 (2016), arXiv:1509.05190.

\bibitem{CMS:2017yer}
CMS, A.~M. Sirunyan {\em et~al.},
\newblock JHEP {\bf 10}, 131 (2017), arXiv:1706.05868.

\bibitem{CMS:2020plq}
CMS, A.~M. Sirunyan {\em et~al.},
\newblock JHEP {\bf 07}, 115 (2020), arXiv:2004.00602.

\bibitem{Chen:2019gqo}
S.-Y. Chen, B.-W. Zhang, and E.-K. Wang,
\newblock Chin. Phys. C {\bf 44}, 024103 (2020), arXiv:1908.01518.

\bibitem{AbdulKhalek:2021gbh}
R.~Abdul~Khalek {\em et~al.},
\newblock Nucl. Phys. A {\bf 1026}, 122447 (2022), arXiv:2103.05419.

\bibitem{Field:1977fa}
R.~D. Field and R.~P. Feynman,
\newblock Nucl. Phys. B {\bf 136}, 1 (1978).

\bibitem{Waalewijn:2012sv}
W.~J. Waalewijn,
\newblock Phys. Rev. D {\bf 86}, 094030 (2012), arXiv:1209.3019.

\bibitem{Liu:2018trl}
X.~Liu, F.~Ringer, W.~Vogelsang, and F.~Yuan,
\newblock Phys. Rev. Lett. {\bf 122}, 192003 (2019), arXiv:1812.08077.

\bibitem{Alrashed:2021csd}
M.~Alrashed, D.~Anderle, Z.-B. Kang, J.~Terry, and H.~Xing,
\newblock Phys. Rev. Lett. {\bf 129}, 242001 (2022), arXiv:2107.12401.

\bibitem{Alrashed:2023xsv}
M.~Alrashed, Z.-B. Kang, J.~Terry, H.~Xing, and C.~Zhang,
\newblock (2023), arXiv:2312.09226.

\bibitem{Miller:2007ri}
M.~L. Miller, K.~Reygers, S.~J. Sanders, and P.~Steinberg,
\newblock Ann. Rev. Nucl. Part. Sci. {\bf 57}, 205 (2007),
  arXiv:nucl-ex/0701025.

\bibitem{Trzcinska:2001sy}
A.~Trzcinska {\em et~al.},
\newblock Phys. Rev. Lett. {\bf 87}, 082501 (2001).

\bibitem{Zenihiro:2010zz}
J.~Zenihiro {\em et~al.},
\newblock Phys. Rev. C {\bf 82}, 044611 (2010).

\bibitem{Sjostrand:2007gs}
T.~Sjostrand, S.~Mrenna, and P.~Z. Skands,
\newblock Comput. Phys. Commun. {\bf 178}, 852 (2008), arXiv:0710.3820.

\bibitem{Dulat:2015mca}
S.~Dulat {\em et~al.},
\newblock Phys. Rev. D {\bf 93}, 033006 (2016), arXiv:1506.07443.

\bibitem{Eskola:2016oht}
K.~J. Eskola, P.~Paakkinen, H.~Paukkunen, and C.~A. Salgado,
\newblock Eur. Phys. J. C {\bf 77}, 163 (2017), arXiv:1612.05741.

\bibitem{Cacciari:2008gp}
M.~Cacciari, G.~P. Salam, and G.~Soyez,
\newblock JHEP {\bf 04}, 063 (2008), arXiv:0802.1189.

\bibitem{Cacciari:2011ma}
M.~Cacciari, G.~P. Salam, and G.~Soyez,
\newblock Eur. Phys. J. C {\bf 72}, 1896 (2012), arXiv:1111.6097.

\bibitem{Kang:2021ryr}
Z.-B. Kang, X.~Liu, S.~Mantry, M.~C. Spraker, and T.~Wilson,
\newblock Phys. Rev. D {\bf 103}, 074028 (2021), arXiv:2101.04304.

\bibitem{Kang:2023ptt}
Z.-B. Kang, A.~J. Larkoski, and J.~Yang,
\newblock Phys. Rev. Lett. {\bf 130}, 151901 (2023), arXiv:2301.09649.

\bibitem{Kang:2013wca}
Z.-B. Kang, X.~Liu, S.~Mantry, and J.-W. Qiu,
\newblock Phys. Rev. D {\bf 88}, 074020 (2013), arXiv:1303.3063.

\bibitem{Anderle:2021wcy}
D.~P. Anderle {\em et~al.},
\newblock Front. Phys. (Beijing) {\bf 16}, 64701 (2021), arXiv:2102.09222.

\bibitem{Guo:1998rd}
X.-f. Guo,
\newblock Phys. Rev. D {\bf 58}, 114033 (1998), arXiv:hep-ph/9804234.

\bibitem{Kang:2013raa}
Z.-B. Kang, E.~Wang, X.-N. Wang, and H.~Xing,
\newblock Phys. Rev. Lett. {\bf 112}, 102001 (2014), arXiv:1310.6759.

\bibitem{Ru:2019qvz}
P.~Ru, Z.-B. Kang, E.~Wang, H.~Xing, and B.-W. Zhang,
\newblock Phys. Rev. D {\bf 103}, L031901 (2021), arXiv:1907.11808.

\bibitem{Ru:2023ars}
P.~Ru, Z.-B. Kang, E.~Wang, H.~Xing, and B.-W. Zhang,
\newblock (2023), arXiv:2302.02329.

\bibitem{Guo:2000nz}
X.-f. Guo and X.-N. Wang,
\newblock Phys. Rev. Lett. {\bf 85}, 3591 (2000), arXiv:hep-ph/0005044.

\bibitem{Wang:2002ri}
E.~Wang and X.-N. Wang,
\newblock Phys. Rev. Lett. {\bf 89}, 162301 (2002), arXiv:hep-ph/0202105.

\bibitem{Li:2020rqj}
H.~T. Li and I.~Vitev,
\newblock Phys. Rev. Lett. {\bf 126}, 252001 (2021), arXiv:2010.05912.

\bibitem{Zhang:2021oki}
S.-L. Zhang, X.-N. Wang, and B.-W. Zhang,
\newblock Phys. Rev. C {\bf 105}, 054902 (2022), arXiv:2103.07836.

\end{thebibliography}

\end{document}